\documentclass[twocolumn,showpacs,superscriptaddress,preprintnumbers,amsmath,amssymb]{revtex4}

\usepackage{epsfig,graphicx,dcolumn,bm,times}
\usepackage{color} 
\usepackage[colorlinks,linkcolor=blue,anchorcolor=green,citecolor=blue]{hyperref}

\begin{document}

\title{Vaccination intervention on epidemic dynamics in networks}

\author{Xiao-Long Peng}
\affiliation{Department of Mathematics, Shanghai University,
Shanghai 200444, China}
\author{Xin-Jian Xu}
\affiliation{Department of Mathematics, Shanghai University,
Shanghai 200444, China}
\affiliation{Institute of Systems Science,
Shanghai University, Shanghai 200444, China}
\author{Xinchu Fu}
\affiliation{Department of Mathematics, Shanghai University,
Shanghai 200444, China}
\affiliation{Institute of Systems Science,
Shanghai University, Shanghai 200444, China}
\author{Tao Zhou}
\affiliation{Web Sciences Center, University of Electronic Science
and Technology of China, Chengdu 610051, China}

\date{\today}

\begin{abstract}
  Vaccination is an important measure available for preventing or reducing the spread of infectious diseases. In this paper, an epidemic model including susceptible, infected and imperfectly vaccinated compartments is studied on Watts-Strogatz small-world, Barab\'{a}si-Albert scale-free, and random scale-free networks. The epidemic threshold and prevalence are analyzed. For small-world networks, the effective vaccination intervention is suggested and its influence on the threshold and prevalence is analyzed. For scale-free networks, the threshold is found to be strongly dependent both on the effective vaccination rate and on the connectivity distribution. Moreover, so long as vaccination is effective, it can linearly decrease the epidemic prevalence in small-world networks, whereas for scale-free networks, it acts exponentially. These results can help adopting pragmatic treatment upon diseases in structured populations.
\end{abstract}


\maketitle

\section{Introduction}

Mathematical characterization of infectious diseases has contributed greatly to getting insight on transmission patterns of a disease in host populations, as well as on public health policies to prevent, reduce, and possibly eradicate the disease \cite{AM91,HWH00,JDM02}. Classical epidemic models usually assume that either individuals do not have immunity to infection (the susceptible-infected-susceptible (SIS) model) or experiencing infection with permanent or temporary protection against it (susceptible-infected-recovered (SIR) and susceptible-infected-recovered-susceptible (SIRS) models). However, there is increasing evidence that most infections, such as pertussis and tuberculosis, can provide only partial immunity and spread among seropositive individuals, regardless of a reduced transmission rate. In view of this fact, vaccination was introduced into mathematical compartmental models which is often represented by a transfer between the susceptible and removed classes \cite{MEH96,KV00,B04,SHD06,LTI08,XT10}. Whether vaccination is inoculation or education, typically it reaches only a fraction of the susceptible populations and is not perfectly effective. Thus, a backward transfer must be considered because vaccinated individuals may return to be susceptible or become directly infected. When these aspects are included in the model, rich dynamical behaviors may arise, such as backward bifurcation and bistability \cite{KV00,B04}.

Previous studies of mathematical models incorporating vaccination either ignore the population structure or treat populations as distributed in a regular medium, that is, all the individuals have the same probability of contacting the others. Recently, classical epidemic models have been extended in many ways (e.g., to study the disease spreading in a population divided into subgroups which may influence each other \cite{NMF03}). Especially, a great source of inspiration to mathematical epidemiology has been provided by the network theory whose nodes represent individuals and links stand for interactions among them \cite{AB02,DM02,MN03,SB06}. The structure of the underlying network (e.g., the degree distribution) may strongly influence spreading dynamics
\cite{LM01,PV01,PV01B,PV02,MN02,LAM03,KE05,CPV07,BBV08,MAM09,PAN09,RD10,EMV11}. For instance, in scale-free (SF) networks, characterized by degree distributions with power-law behavior $P(k) \sim k^{-\gamma}$, the statistical relevance of hubs makes the network highly permeable to disease propagating \cite{PV01}. This radical change in the behavior of the processes suggests that the standard epidemiological frameworks should be carefully revisited.

The mathematical compartmental theory focuses on epidemic equilibria and their stability. The network-based modeling, however, pays much attention to the underlying contact structure among individuals. The goal of this paper is to investigate the influence of vaccination on disease spreading. Different from the classic study of the SIS model with vaccination by the compartmental theory which focuses on the stability of equilibria \cite{KV00}, the present work revisits the model on Watts-Strogatz (WS), Barab\'{a}si-Albert (BA), and random SF networks concentrating on the epidemic threshold and prevalence. The choice of this model is based on three factors as follows. (i) The SIS epidemic framework has been widely used in modeling disease spreading within a population. Each individual is simply assumed to have only one of the two states: susceptible (S) and infected (I). Each susceptible individual gets infection with a transmission rate $\alpha$ once it contacts an infected one. Meanwhile, infected individuals recover and become susceptible again with a recovery rate $\beta$. Then the process of disease transmission flows as S$\rightarrow$I$\rightarrow$S. (ii) Immunization of population through vaccination strategy is an important and feasible practice with obvious implications for the public health. At the population level, it is interesting to determine the critical vaccination rate necessary for eradicating diseases or preventing infection, and to investigate how vaccination affects the epidemic prevalence in the steady state on different networks. In this paper, to study possible effects of vaccination on epidemic dynamics in different networked populations, a vaccinated (V) state is introduced into the SIS model by vaccinating the susceptible individuals with a vaccination rate $\varphi$, corresponding to the transition S$\rightarrow$V. (iii) In the real world, there are various types of vaccines: some may offer temporary immunity; vaccines may not possess 100\% efficacy (leaky vaccines) \cite{MM03}, and finally, on most occasions, vaccination may cover only a fraction of susceptible individuals. Therefore, the vaccinated individuals return to the susceptible class with a resusceptibility rate $\phi$ as the vaccine wears off, or directly get infected with a reduced transmission rate $\delta\alpha$, where $\delta$ denotes the degree to which the vaccine-induced protection against infection is inefficient. Thus, the vaccinated class flows in two directions S$\leftarrow$V$\rightarrow$I.

To study the SIS model with vaccination on networks, both analytical calculations and numerical simulations are carried out. Depending on network structures, two types of epidemic thresholds and corresponding prevalence behavior are obtained. For WS networks, there is a nonzero threshold similar to that obtained in the classic compartmental model. Whereas for SF networks, the threshold is quite different which is strongly related both to the vaccination rate and to the node-degree distribution. While the disease is endemic in the network, it is found that vaccination intervention contributes to linearly reducing the prevalence in WS networks,
whereas in SF networks it functions exponentially.

The rest of the paper is arranged as follows. In Sec. II the SIS model with vaccination is introduced. Then the model is studied on WS, BA, and random SF networks in Secs. III, IV, and V, respectively. Finally, conclusions are given in Sec. VI.

\begin{figure}
  \includegraphics[width=\columnwidth]{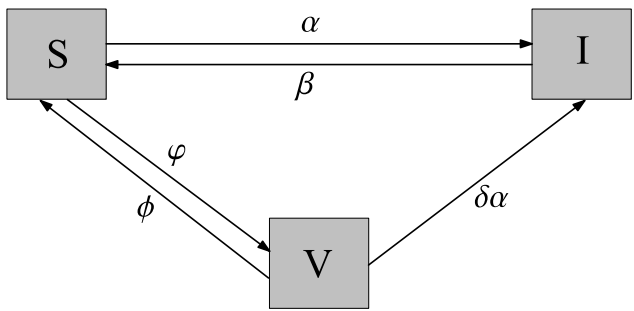}
  \caption{Flowchart of the SIS model with vaccination. Susceptible nodes are infected by their infected neighbors at a per capita rate $\alpha$ and are vaccinated at a per capita rate $\varphi$. Infected nodes recover to be susceptible at a per capita rate $\beta$. Vaccinated nodes become susceptible at a per capita rate $\phi$ and are infected at a per capita rate $\delta\alpha$ due to imperfect vaccination.}\label{fig1}
\end{figure}

\section{The SIS model with vaccination}\label{sivdef}

Vaccination scheme has always been a very important and effective way for preventing or controlling infectious diseases. In reality, vaccines hardly cover the whole population, and only remain effective for a finite period of time, and are difficult to guarantee a perfect protection from infection. Taking all these points into consideration, Kribs-Zaleta and Velasco-Hern\'{a}ndez introduced a vaccinated state into the SIS model to theoretically study the possible effects of vaccination on epidemics \cite{KV00}. The mathematical compartmental model in Ref.~\cite{KV00} neglects the population structure and assumes all individuals have the same contact rate. Therefore, it is intriguing to inspect the vaccination program for different network structures.

In this paper the SIS model with vaccination is formulated on the static network framework, where nodes represent individuals and links stand for the contacts among individuals along which a disease can spread. At each time step, each node exists in only one of the three states: susceptible, infected and vaccinated. A disease
spreads in the network following mechanisms below (as shown in Fig.~\ref{fig1}): a susceptible node will be infected with the transmission rate $\alpha$ once it is connected to an infected one; an infected node is cured and become susceptible again with the recovery rate $\beta$; according to the random vaccination strategy, each susceptible node gets vaccinated with the vaccination rate $\varphi$ and each vaccinated node returns to the susceptible class at the resusceptibility rate $\phi$ as the vaccine wears off; due to the imperfect immunity, a vaccinated node will be infected with a reduced infection rate $\delta\alpha$, where the parameter $\delta$ measures the inefficacy of the vaccine-induced protection against infection ($\delta=0$ and $1$ respectively represent completely effective and utterly invalid, but $0<\delta\ll 1$ holds for most cases \cite{MM03}). In present work it is always assumed that $\delta$ is sufficiently small. The epidemic dynamics is determined by five parameters, $\alpha$, $\beta$, $\varphi$, $\phi$, and $\delta$. For convenience, two ratios $\lambda$, $\eta$ are particularly denoted as $\lambda=\alpha/\beta$ and $\eta=\varphi/\phi$ \cite{Note2}.

In the theoretical study of epidemiology, there are two paramount indicators. One is to formulate the epidemic threshold, which determines whether the infection breaks out in the population and results in an EE or dies out eventually corresponding to a DFE~\cite{Note1}. The other is to predict the epidemic prevalence. The present work make a comprehensive study of the SIS model with vaccination on WS, BA, and random SF networks, employing the mean-field (MF) approach and computational simulations. As will be seen below, vaccination indeed has a great influence on the epidemic dynamics over such networks.

\section{The model on WS networks}\label{sivws}

The WS network \cite{WS98}, as a reference of homogeneous networks, can be constructed as follows. Start from a ring of $N$ nodes, where each node is connected symmetrically with its $2K$ nearest neighbors. Then, every link connected to a clockwise neighbor is rewired to a randomly chosen node with probability $p$. After the whole sweep, a WS network with the average connectivity $\langle k\rangle=2K$ is generated.

The WS network is a typical example of networks characterized by a narrow degree distribution, in which each node's degree closes to $\langle k\rangle$. Let $s(t)$, $\rho(t)$, and $v(t)$ be the densities of susceptible, infected and vaccinated individuals at time $t$, respectively. Obviously, they satisfy the normalization condition $s(t)+\rho(t)+v(t)=1$. Therefore, a set of coupled differential equations can be established following the MF approach \cite{PV01B}:
\begin{widetext}
\begin{subequations}\label{wsmaster} \begin{align}
\frac{\rm d}{{\rm d}t} \rho(t)&=-\beta\rho(t)+\alpha\langle k\rangle s(t)\rho(t)+\delta\alpha\langle k\rangle\big[1-\rho(t)-s(t)\big ] \rho(t),\label{eq:1a}\\
\frac{\rm d}{{\rm d}t} s(t)&=\beta \rho(t)-\alpha\langle k\rangle s(t)\rho(t)+\phi\big[1-\rho(t)-s(t)\big ]-\varphi s(t).\label{eq:1b}
\end{align}
\end{subequations}
\end{widetext}
The first term on the right-hand side (rhs) in Eq.~(\ref{eq:1a}) accounts for the recovery process from the infected class, which is proportional to the recovery rate $\beta$, the average density $\rho(t)$ of infected nodes. The second term on the rhs in Eq.~(\ref{eq:1a}) denotes the newly infected nodes transferred from susceptible ones. It is proportional to the density $s(t)$ of susceptible nodes, the transmission rate $\alpha$, the average number of neighbors $\langle k\rangle$, and the probability $\rho(t)$ that a randomly chosen neighbor is infected. Similarly, the third term on the rhs in Eq.~(\ref{eq:1a}) considers the probability that a node is vaccinated $[1-\rho(t)-s(t)]$, and gets infection. The probability of this last process is proportional to the vaccine-reduced transmission rate $\delta \alpha$, the average number of neighbors $\langle k\rangle$, and the probability $\rho(t)$ that a randomly chosen neighbor is infected. On the rhs in Eq.~(\ref{eq:1b}), the third term accounts for the increment in the susceptible class result from the transition V$\rightarrow$S, which is proportional to the resusceptibility rate $\phi$; and the fourth term accounts for the probability of the vaccination process S$\rightarrow$V, which is proportional to the vaccination rate $\varphi$.

It should be stressed that the MF approach (Eqs.~(\ref{eq:1a}) and (\ref{eq:1b})) is equivalent to the mass-action law (system (1) in Ref.~\cite{KV00}) with the adaptation of the reaction rates to include the average connectivity $\langle k\rangle$. Thus, one can obtain similar results for the epidemic threshold and equilibrium stability. It is due to the homogeneity of the WS network, which is also the case for the SIS model in Ref.~\cite{PV01B}. In accordance with the results in Ref.~\cite{KV00}, there is either a forward bifurcation or a backward one, depending on the epidemiological parameters: (1) in case of the model exhibiting the forward bifurcation, there is only one globally stable EE as the transmission rate is above the epidemic threshold, below which the DFE is the only attractor; (2) in case of the presence of the backward bifurcation, there are multiple endemic equilibria (MEE) - meaning that there are two or more EEs in the steady state - existing between a sub-threshold and the epidemic threshold, meanwhile both the DFE and the lower EE are locally stable.

\begin{figure}
  \begin{center}
  \includegraphics[width=\columnwidth]{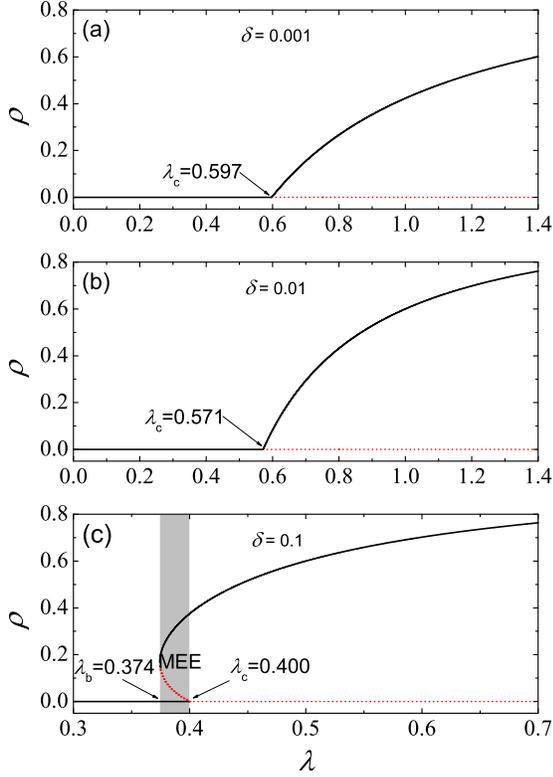}
  \caption{(Color online) Bifurcation diagrams of $\rho$ as a function of $\lambda$ in the WS network with the average connectivity $\langle k\rangle=10$ for various $\delta$: 0.001 (a), 0.01 (b), and 0.1 (c). In each chart, black and red curves respectively represent stable and unstable branches of system~(\ref{wsmaster}). Parameter values: $\beta=0.002$, $\varphi=0.001$, and $\phi=0.0002$.}\label{fig2}
  \end{center}
\end{figure}

In the following much attention will be paid on the epidemic prevalence in the steady state. Imposing the stationary conditions $\frac{\rm d}{{\rm d} t}s(t)=0$ and $\frac{\rm d}{{\rm d}t}\rho(t)=0$ yields
\begin{widetext}
\begin{equation}\label{wsrhofunc} 
\rho =(1-\beta)\rho +\delta\alpha\langle k\rangle(1-\rho)\rho +\alpha\langle k\rangle(1-\delta) \frac{(\beta-\phi)\rho^2+\phi\rho}{\alpha\langle k\rangle\rho+\varphi+\phi} =f(\rho)
\end{equation}
\end{widetext}
for density of infected nodes in steady states. Notice that $f(0)=0$ and $f(1)<1$, Eq.~(\ref{wsrhofunc}) has a nonzero solution on the interval $(0,1)$ only if $f'(\rho)\Big|_{\rho=0}>1$, which defines the epidemic threshold
\begin{equation}\label{wsthreshold}
\lambda_{\rm
c}=\frac{\varphi+\phi}{\delta\varphi+\phi}\frac{1}{\langle k\rangle}=\frac{\eta+1}{\delta\eta+1}\frac{1}{\langle k\rangle}.
\end{equation}
On the other hand, Eq.~(\ref{wsrhofunc}) can be rewritten as 
\begin{equation}\label{wsquadratic}
F(\rho)=A\rho^2+B\rho+C=0,
\end{equation}
with coefficients
\begin{eqnarray}
A &=& \delta\alpha\langle k\rangle, \nonumber\\
B &=& \delta\varphi+\phi+\delta\beta-\delta\alpha\langle k\rangle, \nonumber\\
C &=& -[(\delta\varphi+\phi)-\frac{1}{\lambda\langle k\rangle}(\varphi+\phi)]. \nonumber
\end{eqnarray}
The solutions to Eq.~(\ref{wsquadratic}) correspond to equilibria of system~(\ref{wsmaster}) for given $\lambda$.

\begin{figure}
  \begin{center}
  \includegraphics[width=\columnwidth]{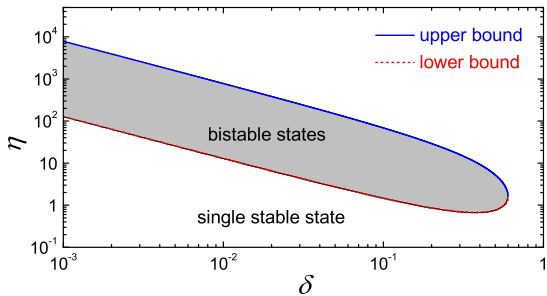}
  \caption{(Color online) Illustration of the dynamic behavior on the $\delta$-$\eta$ plane with parameters $\beta=0.002$ and $\phi=0.0002$, showing the necessary condition for bistability in the WS network with $\langle k\rangle=10$. The white (grey) region corresponds to the single stable state (bistable states). The blue (red) line is upper (lower) bound of $\eta$ for given $\delta$. Only if $\eta$ is inside the range between the upper and lower bounds can the MEE occur. Otherwise, there is only one attractor (either the DFE or the EE).}\label{fig3}
  \end{center}
\end{figure}

(i) $\lambda>\lambda_{\rm c}$. In this case, $C<0$ always holds. Since $F(0)=C<0$ and $F(1)=A+B+C=\delta\beta+(\varphi+\phi)/(\lambda\langle k\rangle)>0$, system (\ref{wsmaster}) has a unique EE,
\begin{equation}\label{wsrhosolution}
\rho = \frac{\sqrt{B^2-4AC}-B}{2A}
\end{equation}
In contrast to special solutions obtained in Ref.~\cite{KV00}, this expression is general. As $\lambda\rightarrow\lambda_{\rm c}$, $C$ closes to $0$. From Eq.~(\ref{wsrhosolution}), it follows that $\rho\rightarrow 0$. Ignoring the second order term of $\rho$ in Eq.~(\ref{wsquadratic}), one has 
\begin{equation}
\rho\approx\frac{\delta\varphi+\phi}{\delta\varphi+\phi+\delta\beta-\delta\alpha\langle k\rangle}\cdot\frac{\lambda-\lambda_{\rm c}}{\lambda_{\rm c}}~~
\sim(\lambda-\lambda_{\rm c}).\label{wsrhosolution2}
\end{equation}
In fact, as given in Appendix A, applying Taylor series expansion to the square root part of the first term at $\delta=0$ and omitting the higher order correction in $\delta$, Eq.~(\ref{wsrhosolution}) can be simplified to
\begin{equation}
\rho\approx 1-\frac{1}{\langle k\rangle}\frac{\eta+1}{\lambda}.\label{wsrho}
\end{equation}

(ii) $\lambda<\lambda_{\rm c}$. In this case, it is impossible to obtain the general solution of the epidemic prevalence. Following the parametric analysis in Ref.~\cite{KV00}, one finds that there are two different EEs in the regime $\lambda_{\rm b}<\lambda<\lambda_{\rm c}$ on the premise of $\lambda_{\rm a}<\lambda_{\rm b}$, where $\lambda_{\rm a}=\frac{\delta(\beta+\varphi)+\phi}{\langle k\rangle \delta\beta}$ corresponds to the condition $B=0$, and $\lambda_{\rm b}=\frac{1}{\langle k\rangle}\bigg(1-\frac{\delta\varphi+\phi}{\delta\beta}+\frac{2} {\delta\beta}\sqrt{\beta\delta(1-\delta)\varphi}\bigg)$ corresponds
to $B^2-4AC=0$. This finding suggests that under the condition $\lambda_{\rm a}<\lambda_{\rm b}$, which is equivalent to
\begin{equation}\label{wsinequality1}
(\delta\varphi+\phi)^2<\beta\delta(1-\delta)\varphi,
\end{equation}
there emerges a sub-threshold $\lambda_{\rm b}$ (the persistence threshold, above which an already established epidemic can persist \cite{GDB06}) and an epidemic threshold $\lambda_{\rm c}$ (the invasion threshold, which still denotes the critical parameter value for invasion of new diseases). In the bifurcation diagram, this sub-threshold corresponds to a saddle-node bifurcation, and the epidemic threshold corresponds to a forward bifurcation, and consequently these two thresholds together exhibit first order transitions between the healthy phase (without disease) and the endemic phase (with disease) (see Fig.~\ref{fig2}(c)). This reveals the hysteresis effect caused by the introduction of vaccination into the infectious disease. However, in other cases, there is only the invasion threshold, and hence only the forward bifurcation (see Figs.~\ref{fig2}(a) and \ref{fig2}(b)).

Clearly, whether the interval $(\lambda_{\rm b}, \lambda_{\rm c})$ is a bistable region or not in the bifurcation diagram of $\rho$ as a function of $\lambda$ is completely determined by the condition~(\ref{wsinequality1}) which can be rewritten as
\begin{equation}\label{wsinequality3}
\delta^2\eta^2+[2-\frac{\beta}{\phi}(1-\delta)]\delta\eta+1<0.
\end{equation}
Only if $\frac{\beta}{\phi}(1-\delta)>4$ can the inequality has solutions on the interval $(\eta_1, \eta_2) \subset (0, \infty)$, which indicates that $\delta_{\rm max}=1-4\frac{\phi}{\beta}$. On the contrary, there is only one single stable state if $\beta/\phi\leq 4$. Figure~\ref{fig3} gives an illustration. In case of $\beta=0.002$ and $\phi=0.0002$, the ratio is $\beta/\phi=10$. To ensure $\frac{\beta}{\phi}(1-\delta)>4$, it demands $\delta<0.6$. In particular, at $\delta=0.001$, the upper and lower bounds of $\eta$ are $127.18(1)$ and $7892.81(6)$, respectively. On the other side, given $\eta=5.0$, only if $0.02(7)<\delta<0.51(1)$ can
system~(\ref{wsmaster}) experience the bistable states. From calculation in Appendix B, $\eta$ reaches the minimum $2/3$ at $\delta=3/8$. The presence of such a bistable region highlights an important but unexpected influence of vaccination on disease spreading. The bifurcation diagrams in Fig. \ref{fig2} correspond to three different values of $\delta$ ($=0.001$, $0.01$, and $0.1$) for $\eta=5.0$. At $\delta=0.1$ the system exhibits the bistable phenomenon, and in order to wipe out the disease, one must ensure that $\lambda<0.37(4)$ rather than $\lambda<0.4$.

\begin{figure}
  \begin{center}
  \includegraphics[width=\columnwidth]{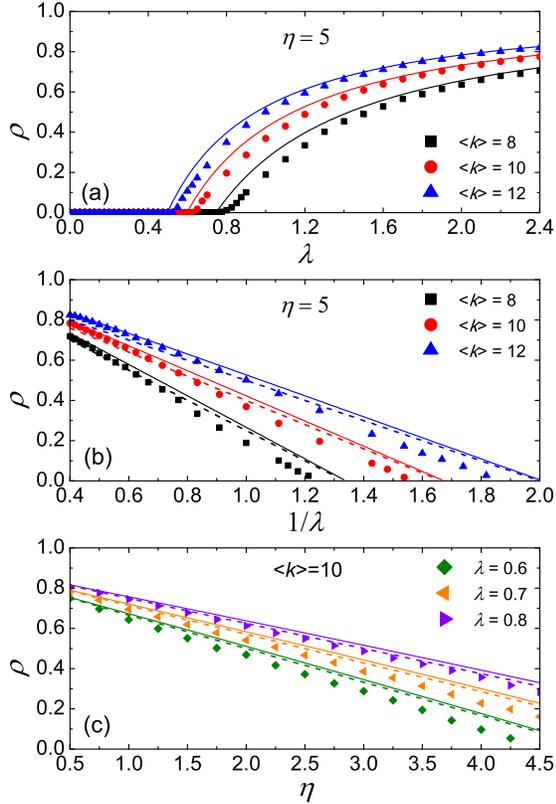}
  \caption{(Color online) Infected densities $\rho$ in the WS networks as functions of $\lambda$ (a),  $1/\lambda$ (b), and $\eta$ (c), respectively. Solid lines are analytical solutions to Eq.~(\ref{wsrhosolution}). Dash lines are theoretical prediction by Eq.~(\ref{wsrho}). Parameters values: $\beta=0.002$, $\phi=0.0002$, and $\delta=0.001$.}\label{fig4}
  \end{center}
\end{figure}

\begin{table*}
    \renewcommand{\arraystretch}{1.2}
    \addtolength{\tabcolsep}{5pt}

  \begin{center}
  \begin{tabular}{|c|c|c|c|c|c|} \hline
   \multicolumn{2}{|c|}{} & \multicolumn{2}{c|}{$D_0$} & \multicolumn{2}{c|}{$D_1$}  \\ \cline{3-6}
   \multicolumn{2}{|c|}{} & {simulation} & analytical   &
   {simulation} & analytical  \\ \hline
    {} & $\langle k\rangle=8$ & 1.07(2)  & 1 & 0.14(5) & 0.125 \\ 
     Fig.~\ref{fig4}(b) & $\langle k\rangle=10$ & 1.06(1) & 1 & 0.11(4) & 0.1 \\ 
    {} & $\langle k\rangle=12$ & 1.06(7) & 1 & 0.09(6) & 1/12 \\ \hline
    {} & $\lambda=0.6$ & 1.01(3) & 1 & 0.11(0) & 0.1 \\ 
     Fig.~\ref{fig4}(c) & $\lambda=0.7$ & 1.00(7) & 1 & 0.10(8) & 0.1 \\ 
    {} & $\lambda=0.8$ & 1.00(3) & 1 & 0.10(4) & 0.1 \\
    \hline
  \end{tabular}
  \caption{Simulation values of $D_0$, $D_1$ calculated by applying the
  Levenberg-Marquardt algorithm \cite{KL44} to the least squares curve fitting on the simulation data plotted in
  Figs.~\ref{fig4}(b) and \ref{fig4}(c), with the general function in the form of
  $\rho=D_0-D_1\frac{\eta+1}{\lambda}$. The quantitative comparison is also
  demonstrated, which shows a good agreement between the
  numerical simulation and the analytical prediction by Eq.~(\ref{wsrho}).}
  \label{tab1}
  \end{center}
\end{table*}

The emergence of MEE gives rise to complexity in the vaccination intervention. In the real world, it is interesting to study the effective vaccination and its influence on the epidemic prevalence. To do that, one can consider system~(\ref{wsmaster}) with proper choice of $\beta$, $\phi$, $\eta$, and $\delta$, respectively, ensuring that only the forward bifurcation occurs. In the following, $\delta$ is fixed at a relatively small value $\delta=0.001$ with $\beta/\phi=10$ and $\eta < 127.18(1)$, where only a globally stable EE exists if $\lambda>\lambda_{\rm c}$ or a globally stable DFE arises if $\lambda \le \lambda_{\rm c}$. According to the model definition, $\lambda$ and $\eta$ are comparably important parameters which affect the global spread of the infection. In Fig.~\ref{fig4} both analytical and numerical results of $\rho$ as a function of $\lambda$ and $\eta$ in the WS network are present, respectively. Simulation of the SIS model with vaccination on the WS network is carried out with parameters $N=10^5$ and $p=0.1$. The fraction of initial infectious seeds is $0.1\%$ and the prevalence $\rho$ in the steady state is averaged over $10$ different realizations of the model on each of $10$ different initial network configurations. Each realization goes through $2\times10^4$ time steps. The thresholds $\lambda_{\rm c}$ in Fig.~\ref{fig4}(a) are $0.77(4)$, $0.62(5)$, and $0.52(3)$, corresponding to the average degrees $\langle k\rangle=8$, $10$, and $12$, respectively, which agrees with the prediction of Eq.~(\ref{wsthreshold}). Moreover, linear behaviors are shown from both the simulation results and the theoretical predictions in Figs.~\ref{fig4}(b) and \ref{fig4}(c).

\begin{figure}
  \begin{center}
  \includegraphics[width=\columnwidth]{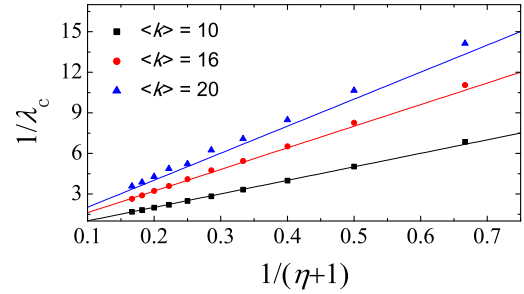}
  \caption{(Color online) $1/\lambda_{\rm c}$ vs $1/(\eta+1)$ in the WS networks for different connectivities. Solid lines correspond to solutions of Eq.~(\ref{wsthreshold}). Parameter values: $\beta=0.002$, $\phi=0.0002$, and $\delta=0.001$.}\label{fig5}
  \end{center}
\end{figure}

To examine the accuracy of analytical prediction by Eq.~(\ref{wsrho}), a quantitative comparison is made in Table~\ref{tab1}, where the Levenberg-Marquardt algorithm is applied to the least squares curve fitting on the simulation data, with the general function in the form of $\rho= D_0-D_1(\eta+1)/\lambda$, where $D_0$ and $D_1$ are positive constants. In Table~\ref{tab1}, the numerical value $D_0$ ranges from 1.00(3) to 1.07(2), matching the theoretical prediction $D_0=1$; and the numerical $D_1$ is also in good agreement with the prediction $D_1=1/\langle k\rangle$. Thus, given all the epidemiological parameters, the larger the average node-degree $\langle k\rangle$, the harder for disease to break out, and if it prevails, the higher the level of infection $\rho$ forms (as shown in Fig.~\ref{fig4}). On the other hand, in a fixed WS network, as the ratio $(\eta+1)/\lambda$ increases, $\rho$ linearly diminishes. This implies that the competition between the transmission process and the vaccination campaign leads to a linear decrease of the prevalence in networks with a narrow degree distribution, as shown in Fig.~\ref{fig4}(c). Notice that for sufficiently small $\delta$, as assumed in present work, Eq.~(\ref{wsrho}) turns into $\rho\approx 1-\lambda_{\rm c}/\lambda$. In form, this scaling behavior of $\rho$ resembles that in the SIS model without vaccination \cite{PV01B}, where the threshold is $\lambda_{\rm c}=1/\langle k\rangle$ . This similarity,
however, reveals that the linear effect of vaccination on the prevalence is in essence due to the fact that vaccination program increases the epidemic threshold by $\eta$ times. To get further information, the inverse of $\lambda_{\rm c}$ as a function of the inverse of $\eta+1$ is depicted in Fig.~\ref{fig5}. Since Eq.~(\ref{wsthreshold}) can be rewritten as $1/\lambda_{\rm c}=\langle k\rangle [(1-\delta)/(\eta+1)+\delta]$, simulation results verify this linearity. Hence, the more effectively the vaccination intervenes on the disease, the more difficultly it outbreaks.

\section{The model on BA networks}\label{sivba}

The BA network \cite{BA99}, as a prototype of heterogeneous networks, can be built as follows. Start from a a set of $m_0$ nodes, which are completely connected. At each time step, a new node is added to the existing network, bringing $m(\leq m_0)$ new links connecting to old nodes with degree preference. After iterating this procedure a sufficient number of times, a BA network is obtained, consisting of $N$ nodes with the node-degree distribution $P(k)=2m^2k^{-3}$ and the mean node-degree $\langle k\rangle=2m$.

The heterogeneity of the connectivity distribution inherent to BA networks induces strong fluctuations, so systems~(\ref{wsmaster}) should be modified accordingly. Denoting by $s_k(t)$, $\rho_k(t)$, and $v_k(t)$ the relative densities of susceptible, infected and vaccinated nodes with degree $k$ at time $t$, respectively, which satisfy the normalization condition $s_k(t)+\rho_k(t)+v_k(t)=1$, the MF equations now read as:
\begin{widetext}
\begin{subequations}\label{bamaster}\begin{align}
\frac{\partial}{\partial t} \rho_k(t)&=-\beta\rho_k(t)+\alpha ks_k(t)\Theta(\rho(t))+\delta\alpha k\big[1-\rho_k(t)-s_k(t)\big]\Theta(\rho(t)),\label{eq:10a}\\
\frac{\partial}{\partial t} s_k(t) &= \beta\rho_k(t)-\alpha ks_k(t)\Theta(\rho(t))+\phi\big[1-\rho_k(t)-s_k(t)\big]-\varphi s_k(t).\label{eq:10b}
\end{align}
\end{subequations}
\end{widetext}
The first term on the rhs in Eq.~(\ref{eq:10a}) considers that a node of degree $k$ is in the infected state with probability $\rho_k(t)$ and recovers from infection at the recovery rate $\beta$. The second term on the rhs in Eq.~(\ref{eq:10a}) considers the probability that a node with $k$ links is in the susceptible state $s_k(t)$ and gets infection via a neighbor. The probability of this last event is proportional to the transmission rate $\alpha$, the number of neighbors $k$, and the probability $\Theta(\rho(t))$ that any given link points to an infected node. Similarly, the third term on the rhs in Eq.~(\ref{eq:10a}) considers that a node with k neighbors is in the vaccinated state with probability $[1-\rho_k(t)-s_k(t)]$ and gets infection via a neighbor at the vaccine-reduced transmission rate $\delta\alpha$. On the rhs in Eq.~(\ref{eq:10b}), the third term considers that a node with degree $k$ is in the vaccinated state with probability $[1-\rho_k(t)-s_k(t)]$ and returns to the susceptible class at the resusceptibility rate $\phi$; and the fourth term considers that a node of degree $k$ is susceptible with probability $s_k(t)$ and gets vaccinated at the vaccination rate $\varphi$. For uncorrelated networks, the probability $\Theta$ is \cite{PV01B}
\begin{equation}\label{batheta1}
\Theta=\sum_{k} \frac{kP(k)}{\sum_{s} sP(s)}\rho_{k}.
\end{equation}
Since SF networks have no correlations under the constraint that the maximum possible degree has a cutoff scaling at most as $k_{\rm c}(N)\sim N^{1/2}$ \cite{CBP05}. In order to ensure an uncorrelated BA network, this restriction on the maximum degree is imposed in present work. Imposing the stationary conditions $\frac{\partial}{\partial t}\rho_k(t)=0$ and $\frac{\partial}{\partial t}s_k(t)=0$ yields
\begin{equation}\label{barhok}
\rho_k=\frac{\alpha k\Theta\big(\delta\varphi+\phi+\delta\alpha k\Theta\big)} {\alpha k\Theta\big(\delta\varphi+\phi+\delta\alpha k\Theta\big) +\beta\big(\varphi+\phi+\delta \alpha k\Theta\big)}.
\end{equation}
Combining Eqs.~(\ref{batheta1}) and (\ref{barhok}), one obtains a self-consistency equation,
\begin{eqnarray}\label{batheta2}
\Theta &=& \frac{1}{\langle k\rangle}\sum_k\frac{kP(k)\alpha k\Theta\big(\delta\varphi+\phi+\delta\alpha k\Theta\big)} {\alpha
k\Theta\big(\delta\varphi+\phi+\delta\alpha k\Theta\big) +\beta\big(\varphi+\phi+\delta \alpha k\Theta\big)} \nonumber\\
&=& g(\Theta).
\end{eqnarray}
Obviously, there is a trivial solution $\Theta=0$ which leads to $\rho=0$. Notice that not only $g(0)=0$ and $g(1)<1$, but also $g'(\Theta)>0$, and $g''(\Theta)<0$ in the limit $\delta\rightarrow 0$, only if $g'(\Theta)\Big |_{\Theta=0}>1$ can Eq.~(\ref{batheta2}) have a nontrivial solution on the interval $(0,1)$, which yields \begin{equation}\label{balambdac}
\lambda_{\rm c} =\frac{\varphi+\phi}{\delta\varphi+\phi}\frac{\langle k\rangle}{\langle k^2\rangle} =\frac{\eta+1}{\delta\eta+1}\frac{\langle k\rangle}{\langle k^2\rangle}.
\end{equation}
In infinite-sized BA networks, the second moment of the connectivity distribution is unbounded, i.e., $\langle k^2 \rangle \rightarrow \infty$, which induces $\lambda_{\rm c}=0$. So the infection can always prevail among the population, no matter what the effective transmission rate is. Whereas for finite-sized BA networks, there exists a maximum degree $k_{\rm c}$, which controls the bound of the connectivity fluctuations, inducing a nonzero threshold \cite{LM01,PV02}. From now on, the size of the BA networks is assumed to be finite, and all the possible values of node degrees are $k=m, m+1, \ldots, k_{\rm c}$.

By computing the Jacobian matrix of the DFE $\{(\rho_k=0, s_k=\frac{\phi}{\varphi+\phi})\}_{k=m}^{k_{\rm c}}$ of system (\ref{bamaster}), one finds that the basic reproductive number \cite{AM91} is $R_0=\lambda\frac{\delta\eta+1}{\eta+1}\frac{\langle k^2\rangle}{\langle k\rangle}=\lambda/\lambda_{\rm c}$, which denotes the expected number of secondary infections caused by a single infected individual in a completely susceptible population. Accordingly, the DFE is locally asymptotically stable if $\lambda<\lambda_{\rm c}$, while unstable if $\lambda>\lambda_{\rm c}$ meaning invasion is always possible. As long as $\lambda>\lambda_{\rm c}$ there exists a positive solution $\rho\in(0,1)$ corresponding to the EE which is locally asymptotically stable.

Since $\Theta$ approaches 0 as $\lambda$ closes to $\lambda_{\rm c}$, and by neglecting all higher order corrections in $\Theta$, Eq.~(\ref{barhok}) is in form analogous to Eq.~(8) in Ref.~\cite{PV01B}, and hence one expects the similar critical behavior given by Eq.~(\ref{balambdac}). Compared with the SIS model on BA networks \cite{PV01,PV01B}, the presence of vaccination has the effect of multiplying the epidemic threshold by a factor $(\eta+1)/(\delta\eta+1)$, i.e., enlarging by nearly $\eta$ times (as $\delta\rightarrow 0$). This suggests that vaccination might play a significant role in preventing or reducing the infectious disease. The greater the vaccination rate is, the bigger the epidemic threshold is, and hence the harder the disease erupts. 

Neglecting the second order term in $\Theta$, Eq.~(\ref{barhok}) can be simplified as
\begin{equation}
\rho_k\approx \frac{\lambda \Theta k}{\lambda \Theta k+\eta+1}.
\end{equation}
Given the epidemiological parameters, a nodes with higher degree is more likely to get infected. Substituting this expression into Eq.~(\ref{batheta2}) and treating $k$ as a continuous variable yields
\begin{equation}
\Theta\approx m\lambda\Theta\int_{m}^{k_{\rm c}}\frac{1}{k}\frac{{\rm d}k}{\lambda\Theta k+\eta+1},
\end{equation}
which gives rise to the solution
\begin{equation}\label{bathetaapprox}
\Theta\approx \frac{(\eta+1)e^{-(\eta+1)/{\lambda m}}}{\lambda m}\big[1-e^{-(\eta+1)/{\lambda m}}\big]^{-1}.
\end{equation}
Finally, at lowest order in $\lambda$, the epidemic prevalence related to the EE is
\begin{equation}\label{barhoapprox}
\rho = \sum_k P(k)\rho_k \approx 2e^{-(\eta+1)/{\lambda m}}.
\end{equation}

\begin{figure}
  \begin{center}
  \includegraphics[width=\columnwidth]{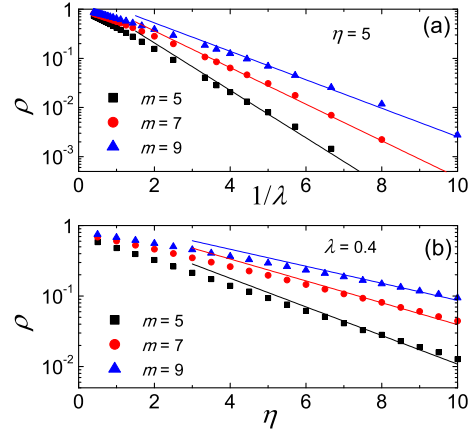}
  \caption{(Color online) Densities of infectious nodes $\rho$ in the BA networks: (a) as a function of $1/\lambda$ for $\eta=5.0$ and various $\langle k\rangle$; (b) as a function of $\eta$ for $\lambda=0.4$ and various $m$. Solid lines are theoretical prediction by Eq.~(\ref{barhoapprox}). The other parameters are $\beta=0.002$, $\phi=0.0002$, and $\delta=0.001$.}\label{fig6}
  \end{center}
\end{figure}

\begin{table}
  \renewcommand{\arraystretch}{1.1}
  \addtolength{\tabcolsep}{5pt}
  \begin{center}
  \begin{tabular}{|c|c|c|c|}
  \hline
  \multicolumn{2}{|c|}{} & \multicolumn{2}{c|}{$E_0$}  \\ \cline{3-4}
  \multicolumn{2}{|c|}{} & {simulation} & analytical   \\
  \hline
    {} & $m=5$ & 0.18(4) & 0.2   \\ 
     Fig.~\ref{fig6}(a) & $m=7$ & 0.12(7) & 1/7  \\ 
    {} & $m=9$ & 0.09(8) & 1/9   \\ \hline
    {} & $m=5$ & 0.17(7) & 0.2  \\ 
     Fig.~\ref{fig6}(b) & $m=7$ & 0.11(9) & 1/7  \\ 
    {} & $m=9$ & 0.09(3) & 1/9  \\
  \hline
  \end{tabular}
  \caption{Simulation value of $E_0$ calculated by applying the Levenberg-Marquardt algorithm to the least squares curve fitting on the simulation data plotted in Figs.~\ref{fig6}(a) and \ref{fig6}(b). The quantitative comparison is also demonstrated, between the fitting value $E_0$ and the analytical prediction $1/m$ by Eq.~(\ref{wsrho}).}
  \label{tab2}
  \end{center}
\end{table}

Computational simulations for the epidemic model are performed on the BA networks with the network size $N=10^5$. Each of the simulation data is obtained by averaging over $10$ different realizations of the model on each of $10$ different network configurations. Each realization goes through $2\times10^4$ time steps. As shown in Fig.~\ref{fig6}, there is a deviation between the simulation results and the analytical calculations, especially in the large prevalence regime. It is due to the fact that Eq.~(\ref{barhok}) is simplified by neglecting the highest order in $\Theta$. As $\rho$ is relatively large, $\Theta^2$ is actually not negligible. Despite this, the simulation support the calculation by the same exponential decaying in the scaling behavior, i.e., $\rho\sim e^{-E_0(\eta+1)/\lambda}$, where $E_0$ is a constant. The numerical comparison is also made between the fitting value $E_0$ and the analytical estimation $1/m$ in Table~\ref{tab2}, showing a relatively small variance. For BA networks, both in simulation and in theory, the prevalence decays exponentially, i.e., $\rho\sim e^{-(\eta+1)/\lambda m}$. Vaccination has an effect of accelerating by $\eta$ times the exponential decreasing of the prevalence. This finding suggests that the vaccination intervention on a disease can efficiently reduce an endemic to a lower level, though the heterogeneity in degree distribution causes a vulnerability to disease outbreak in BA networks.

\section{The model on random SF networks}\label{sivrsf}

In this section, the analysis for the SIS model with vaccination on BA networks will be generalized to random SF networks with arbitrary exponent $\gamma>2$. Following the idea proposed by Newman et al. \cite{NSW01}, the random SF networks can be generated as below. First, a priori random integers sequence, each of which represents the degree of a node, drawn from a normalized distribution
\begin{equation}
P(k)=\left\{
\begin{array}{ccc}
(\gamma-1)m^{\gamma-1}k^{-\gamma} && \quad {\rm if} \quad k\leq k_{\rm c}, \\
0 && \quad {\rm otherwise},
\end{array}\right.
\end{equation}
where $m$ and $k_{\rm c}$ are respectively assumed to be the minimum and the maximum values of the degree among all the nodes, and $k_{\rm c} \gg m$. Notice that in order to get uncorrelated random SF networks, the restriction on the maximum degree \cite{CBP05} $k_{\rm c}(N)\sim N^{1/2}$ is imposed. Then, node $i$ with degree $k_i$ is picked out randomly from the sequence and connected to others until its degree quota $k_i$ is realized. Duplicate connections are avoided. This process is repeated throughout all the elements of the sequence, and finally a network is chosen uniformly at random from the set of all graphs with that degree sequence.
Assuming $k$ changes continuously and the average connectivity is thus
\begin{equation}\label{rsfk}
\langle k\rangle = \int_m^{k_{\rm c}}kP(k){\rm d}k \approx \frac{\gamma-1}{\gamma-2}m.
\end{equation}

For any connectivity distribution in random SF networks, one can employ directly the analytical treatment in BA networks. That is to say, the MF results in the BA network are applicable to the random SF network. According to Eq.~(\ref{balambdac}), the epidemic threshold is zero if $\gamma \le 3$ in the thermodynamic limit for random SF networks. Whereas for $\gamma > 3$, substituting Eq.~(\ref{rsfk}) into Eq.~(\ref{balambdac}) yields a non-zero threshold
\begin{equation}\label{rsflambdac}
\lambda_{\rm c} \approx \frac{(\eta+1)(\gamma-3)}{m(\delta\eta+1)(\gamma-2)}.
\end{equation}
Since $\frac{\partial}{\partial\eta}\lambda_{\rm c} = \frac{(1-\delta)(\gamma-3)}{(\delta\eta+1)^2(\gamma-2)m} < 0$ for any $\delta \in (0,1)$, enhancing the effective vaccination rate can prevent epidemics from spreading through the population. For any exponent $\gamma$ in random SF networks, combining Eqs.~(\ref{batheta2}) and (\ref{rsfk}), one has
\begin{equation}\label{rsfinte1}
\Theta=\frac{\lambda'\Theta(\gamma-1)m^{\gamma-1}}{\langle k\rangle}\int_m^{k_{\rm c}}\frac{k^{2-\gamma}}{\lambda'\Theta k+1}{\rm d}k,
\end{equation}
with
\begin{equation}\label{rsflamb}
\lambda'=\frac{\lambda}{\eta+1}.
\end{equation}
Due to the existence of the parameter $\gamma$, it is difficult to obtain the explicit solution of Eq.~(\ref{rsfinte1}). However, one can roughly estimates $\Theta$ and $\rho$ using the first mean value theorem. In this way, one has
\begin{equation}
\Theta=\frac{\lambda'\Theta(\gamma-2)m^{\gamma-2}}{1+\lambda'\Theta\Omega_1} \int_m^{k_{\rm c}}k^{2-\gamma}{\rm d}k,
\end{equation}
where $\Omega_1$ is a finite constant, $m<\Omega_1<k_{\rm c}$. Thus, the solution is
\begin{equation}\label{rsfthetaapprox}
\Theta\approx\left\{
\begin{array}{ccr}
\frac{1}{\lambda'\Omega_1}\left[\frac{\lambda'(\gamma-2)m^{\gamma-2}k_{\rm c}^{3-\gamma}}{3-\gamma}-1\right] &{\rm if}& 2 < \gamma \le 3,\\
\frac{1}{\lambda'\Omega_1}\left[\frac{\lambda'(\gamma-2)m}{\gamma-3}-1\right] &{\rm if}& \gamma > 3.
\end{array}\right.
\end{equation}
The prevalence $\rho$ can also be written as
\begin{equation}\label{rsfrho}
\rho = \int_m^{k_{\rm c}}P(k)\rho_k{\rm d}k \approx \frac{m(\gamma-1)}{\Omega_2(\gamma-2)}(1-\frac{1}{1+\lambda'\Theta\Omega_2}),
\end{equation}
where $\Omega_2$ is a finite constant, $m<\Omega_2<k_{\rm c}$. According to Eq.~(\ref{rsfthetaapprox}), the behavior of $\rho$ depends on $\gamma$.

(i) $2 < \gamma \le 3$. In this case, for any $\lambda \gg 0$, $\lambda'\Theta\Omega_2\gg 1$. Implementing logarithm operation on Eq.~(\ref{rsfrho}) yields
\begin{equation}
\ln\rho\approx \ln\frac{m(\gamma-1)}{\Omega_2(\gamma-2)}-\frac{1}{1+\lambda'\Theta\Omega_2}.
\end{equation}
Combining this with Eq.~(\ref{rsflamb}), one has
\begin{equation}\label{rsfrho1}
\rho\sim e^{-v_1(\eta+1)/\lambda},
\end{equation}
where $v_1$ is a constant, defined by
\begin{equation}\label{v1}
v_1=\frac{\Omega_1(3-\gamma)}{\Omega_2m^{\gamma-2}k_{\rm c}^{3-\gamma}(\gamma-2)}.
\end{equation}

(ii) $\gamma>3$ while $\gamma\not\gg 3$. According to Eqs.~(\ref{rsflambdac}) and (\ref{rsfthetaapprox}), for any $\lambda \gg \lambda_{\rm c}$, $\lambda'\Theta\Omega_2 \gg 0$. One can obtain the prevalence similar to case (i)
\begin{equation}\label{rsfrho2}
\rho\sim e^{-v_2(\eta+1)/\lambda},
\end{equation}
where the coefficient $v_2$ reads as
\begin{equation}\label{v2}
v_2=\frac{\Omega_1(\gamma-3)}{\Omega_2m(\gamma-2)}.
\end{equation}

(iii) $\gamma \gg 3$. In this case, the connectivity distribution decays so fast that it tends to a homogeneous networks. One would expect to obtain the similar qualitative behavior as in Sec.~\ref{sivws}.

\begin{figure}
  \begin{center}
  \includegraphics[width=\columnwidth]{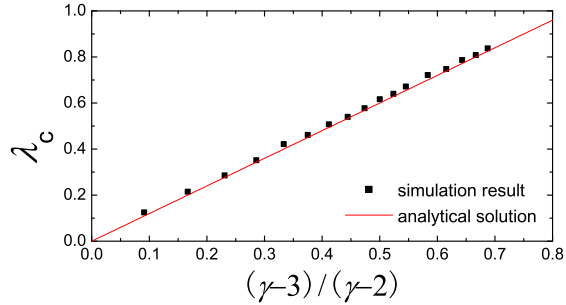}
  \caption{(Color online) Effective transmission threshold $\lambda_{\rm c}$ as a function of $(\gamma-3)/(\gamma-2)$ in the random SF network. The full line corresponds to the analytical calculation of Eq.~(\ref{rsflambdac}). Parameter values: $\eta=5.0$, $\beta=0.002$, $\phi=0.0002$, and $\delta=0.001$.}\label{fig7}
  \end{center}
\end{figure}

\begin{figure}
  \begin{center}
  \includegraphics[width=\columnwidth]{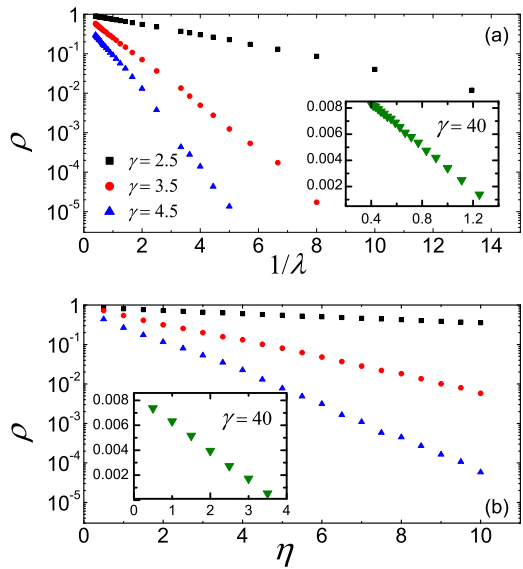}
  \caption{(Color online) Semi-log plots of the persistence $\rho$ in random SF networks as a function of $1/\lambda$ (with $\eta=5.0$) (a) and $\eta$ (with $\lambda=0.5$) (b) for various values of $\gamma$: $2.5$, $3.5$, and $4.5$ (from top to bottom). The insets of (a) and (b) respectively display the linear dependence of $\rho$ as a function of $1/\lambda$ and $\eta$ for $\gamma=40$. Parameter values: $\beta=0.002$, $\phi=0.0002$, and $\delta=0.001$.}\label{fig8}
  \end{center}
\end{figure}

Simulations of the SIS model with vaccination on random SF networks are performed to compare with the theoretical analysis. The simulated networks range from $N=10^5$ to $N=10^6$ and the minimal degree of nodes is $m=5$. Figure~\ref{fig7} shows the epidemic threshold $\lambda_{\rm c}$ as a function of the algebraic expression $(\gamma-3)/(\gamma-2)$. Closed squares represent numerical data and the solid line corresponds to the prediction of Eq.~(\ref{rsflambdac}). One notices the good agreement between the computer simulation and the analytical calculation. Figure~\ref{fig8} depicts the behaviors of $\rho$ as a function of $1/\lambda$ (Fig.~\ref{fig8}(a)) and as a function of $\eta$ (Fig.~\ref{fig8}(b)), respectively. It is clear that either for $\lambda\gg 0$ (the case of $2<\gamma\leq 3$) or for $\lambda \gg \lambda_{\rm c}$ (the case of $\gamma>3$ and $\gamma\not\gg 3$), the stationary density $\rho$ of infected nodes in the random SF networks decays exponentially, i.e., $\rho\sim e^{-{v}(\eta+1)/\lambda}$, where $v$ is a positive constant, which is determined by Eq.~(\ref{v1}) or Eq.~(\ref{v2}). On the contrary,
at $\gamma = 40$, as shown in the inset of Fig.~\ref{fig8}, $\rho$ decreases linearly similar to the behavior observed from WS networks.

\section{Conclusion}

The study of vaccination in populations has to take into consideration not only vaccine-related parameters, but also social risk behaviors that may alter the expected predictions. To our knowledge, however, very few work addressed this problem. The present research integrated the both factors and studied a networked
SIS model with vaccination, where vaccines that attempt to reduce susceptibility to infection is characterized by three parameters in the model: coverage (represented by $\varphi$), waning period (represented by $\phi$), and efficacy (represented by $\delta$). Since $\delta$ is intrinsically related to the quality of the vaccine, much attention has been paid to the parameter $\eta$ (the ratio of $\varphi$ to $\phi$) in the vaccination intervention on infectious diseases, as well as the role of the ratio $\lambda$ (the ration of $\alpha$ to $\beta$) in epidemic spreading. With the frameworks of the MF approach and elementary means, the model has been studied on WS, BA, and random SF networks. The analysis of thresholds and prevalence demonstrated the significant effects of the vaccination on the epidemic dynamics as well as the structures of the underlying networks.

In the WS networks, since the MF model is equivalent to the classic compartmental model in Ref.~\cite{KV00} with the adaptation of reaction rates by the average connectivity, the threshold behavior and equilibrium stability are similar to the literature. The threshold $\lambda_{\rm c}$ is defined by Eq.~(\ref{wsthreshold}), above which there is only one globally stable EE and below which the model may exhibit MEE for certain epidemiological parameters. As to the prevalence, rather than special solutions obtained in the compartmental model, this paper gives the general one for the steady endemic state, which scales as $\rho \sim -(\eta+1)/\lambda$.
Thus, the effective vaccination can linearly decrease the endemic level in homogeneous networks, although vaccination intervention may give rise to the backward bifurcation in these networks.

In the SF networks, however, the system shows very different behavior. The threshold $\lambda_{\rm c}$ is defined by Eq.~(\ref{balambdac}). Only for the SF network with the power-law distribution exponent $2<\gamma\leq 3$ in the thermodynamic limit can $\lambda_{\rm c}$ be zero. Otherwise, the system has a non-zero threshold for the SF networks with any $\gamma>3$. In comparison with the WS network at the same average connectivity $\langle k \rangle$, $\lambda_{\rm c}$ in the SF network is smaller than that in the WS network. For any $\lambda > \lambda_{\rm c}$, the prevalence in the SF network scales as $\rho \sim e^{-v(\eta+1) /\lambda}$. Thus, the vaccination can exponentially decrease the endemic level in heterogeneous networks.

All these results are on the presumption that the underlying networks are static. For some diseases which spread too fast in comparison with change of the population structure, the present work may provide a preliminary theory for vaccine control of infection. For other diseases, however, individual responses to infection plays
an important role in either reducing the transmission rate or changing the contact structure \cite{GB08,SS10,FSJ10}. Hence, it is interesting to study vaccination in adaptive networks, which is left for future research.

\section*{Acknowledgments}

This work was supported by Natural Science Foundation of China (10805033, 10975126, and 11072136), Shanghai Municipal Education Commission (13YZ007), and Shanghai University Leading Academic Discipline Project (A.13-0101-12-004). The authors acknowledge referees for their insightful suggestions.

\section*{Appendix A: Simplification of Eq.~(\ref{wsrhosolution}) to Eq.~(\ref{wsrho})}
\setcounter{equation}{0}
\renewcommand{\theequation}{A\arabic{equation}}

Since the inefficacy rate of vaccine $\delta$ is assumed to be sufficiently small, it is possible to simplify the complex square root part in Eq.~(\ref{wsrhosolution}) via Taylor series expansion at the point $\delta=0$. First of all, rewrite Eq.~(\ref{wsrhosolution}) as
\begin{equation}\label{ap1}
\rho=\rho_1+\rho_2,
\end{equation}
where
\begin{equation}\label{ap2}
\rho_1=\frac{\langle k\rangle\alpha\delta-(\varphi\delta+\beta\delta+\phi)}{2\langle k\rangle\alpha\delta},
\end{equation}
and
\begin{equation}\label{ap3}
\rho_2=\frac{h(\delta)}{2\langle k\rangle\alpha\delta}
\end{equation}
accompanied with
\begin{equation}\label{sqrt}
h(\delta)=\big[h_1(\delta)\big]^{1/2},
\end{equation}
\begin{equation}\label{h1}
h_1(\delta)=\big(\delta\varphi+\phi+\delta\beta-\delta\alpha\langle k\rangle\big)^2+4\delta\alpha\langle k\rangle(\delta\varphi+\phi)-4\delta\beta (\varphi+\phi).
\end{equation}
So, rearranging each term, one has
\begin{equation}
h_1(\delta)=a\delta^2+b\delta+\phi^2,
\end{equation}
where
\begin{equation}\label{ab}
\left\{\begin{array}{ccc} 
a &=& \big(\varphi+\beta-\langle k\rangle \alpha\big)^2+4\langle k\rangle \alpha\varphi,\\
b &=& 2\langle k\rangle\alpha\phi+2\varphi\phi-2\beta\phi-4\beta\varphi.
\end{array}\right.
\end{equation}
Since the first order derivative of $h(\delta)$ can be calculated as
\begin{equation}\label{deri}
 h'(\delta) = \frac{1}{2}\big[h_1(\delta)\big]^{-1/2}(2a\delta+b),
\end{equation}
one has
\begin{equation}
h'(0)=\frac{\langle k\rangle\alpha\phi+\varphi\phi-\beta\phi-2\beta\varphi}{\phi}.
\end{equation}
Therefore, by employing Taylor series expansion at $\delta=0$ for $h(\delta)$ with regard to $\delta$, one gets
\begin{eqnarray}\label{taylor}
h(\delta)&=&h(0)+h'(0)\delta+\circ(\delta^2) \nonumber\\
&\approx &\phi+\frac{\langle k\rangle\alpha\phi+\varphi\phi-\beta\phi-2\beta\varphi}{\phi}\delta.
\end{eqnarray}
Combining Eq.~(\ref{taylor}) with Eqs.~(\ref{ap1}, \ref{ap2}, \ref{ap3}) gives rise to
\begin{equation}
\rho\approx \frac{2\langle k\rangle\alpha\delta-2\beta\delta-2\beta\frac{\varphi}{\phi}\delta}{2\langle k\rangle\alpha\delta},
\end{equation}
which implies the simple relationship
\begin{equation}
\rho\approx 1-\frac{1}{\langle k\rangle}\frac{\eta+1}{\lambda}.
\end{equation}

\section*{Appendix B: Calculation of the minimum lower bound of $\eta$}
\setcounter{equation}{0}
\renewcommand{\theequation}{B\arabic{equation}}

Let $x=\beta/\phi$ and substitute it into inequality (\ref{wsinequality3}), one obtains
\begin{equation}\label{Aineq1}
\delta^2\eta^2+[2-x(1-\delta)]\delta\eta+1<0,
\end{equation}
which has positive solutions if and only if
\begin{equation}\label{Aineq2}
\{[x(1-\delta)-2]^2-4\}\delta^2>0,~~~\Longleftrightarrow~~~~x(1-\delta)>4.
\end{equation}
The solutions of inequality~(\ref{Aineq1}) read
\begin{equation}
\eta_1(\delta)<\eta<\eta_2(\delta),
\end{equation}
where
\begin{equation}\label{Aeta1}
\eta_1(\delta)=\frac{[x(1-\delta)-2]-\sqrt{[x(1-\delta)-2]^2-4}}{2\delta},
\end{equation}
\begin{equation}
\eta_2(\delta)=\frac{[x(1-\delta)-2]+\sqrt{[x(1-\delta)-2]^2-4}}{2\delta}.
\end{equation}
The derivative of the lower bound $\eta_1(\delta)$ is
\begin{equation}
\frac{{\rm d}\eta_1}{{\rm d}\delta} = \frac{1}{2\delta^2}\Big\{\frac{x(x-4)-x(x-2)\delta} {\sqrt{[x(1-\delta)-2]^2-4}} -(x-2)\Big\}.
\end{equation}
Let $\frac{{\rm d}\eta_1}{{\rm d}\delta}=0$, it follows that
\begin{equation}\label{Aeq1}
x[x-4-(x-2)\delta]=(x-2)\sqrt{[x(1-\delta)-2]^2-4},
\end{equation}
which gives the extreme point
\begin{equation}
\delta^*=\frac{x-4}{2(x-2)}=\frac{\frac{\beta}{\phi}-4}{2(\frac{\beta}{\phi}-2)}.
\end{equation}
Substituting $\delta^*$ into Eq.~(\ref{Aeta1}) yields the minimal lower bound
\begin{equation}
{\eta_1}_{\rm min}=\frac{4}{x-4}=\frac{4}{\frac{\beta}{\phi}-4}.
\end{equation}


\end{document}